\newcommand{\ra}{\rangle}

\newcommand{\la}{\langle}

\newcommand{\Dl}{\Delta \lambda}
\newcommand{\lam}{\lambda}

\newcommand{\tilpar}{\tilde{\partial}}
\newcommand{\tilk}{\tilde{k}}
\newcommand{\tildelta}{\tilde{\delta}}
\documentclass[12pt]{article}

\usepackage{latexsym}

\begin{document}

\title{Large-distance behaviour of the graviton 
two-point function in de Sitter spacetime}

\author{Atsushi Higuchi$^1$ and Spyros S.\ Kouris$^2$\\ 
{\normalsize Department of Mathematics, University of York}\\ 
{\normalsize Heslington, York, YO10 5DD, United Kingdom}\\
{\normalsize $^1$Email: ah28@york.ac.uk}\\
{\normalsize $^2$Email: ssk101@york.ac.uk}}

\date{7 July, 2000}
\maketitle

\begin{abstract}
It is known that the graviton two-point function for the de\ Sitter invariant
``Euclidean" vacuum in a physical gauge grows
logarithmically with distance in spatially-flat de\ Sitter spacetime. 
We show that this logarithmic behaviour 
is a gauge artifact by explicitly demonstrating that the same behaviour can
be reproduced by a pure-gauge two-point function. 
\end{abstract}

\section{Introduction}

De\ Sitter spacetime is a maximally symmetric 
solution of the vacuum Einstein equations with 
positive cosmological constant,
\begin{equation}
R_{ab} -\frac{1}{2}g_{ab}R + \Lambda g_{ab} = 0\,. \label{ES}
\end{equation}
(See Ref.\ \cite{HawkEllis} for a detailed description of de\ Sitter
spacetime.)
Physics in this spacetime has been studied
extensively due to its 
relevance to inflationary cosmologies \cite{Inflation}.
The graviton two-point function 
has been of particular interest in this context.
Ford and Parker analysed linearised gravity in spatially-flat
de\ Sitter spacetime and
found that the mode functions in a physical gauge are similar to those of
minimally-coupled massless scalar field \cite{FordParker}.
Since the latter theory exhibits infrared (IR) divergences similar to those
of massless scalar field theory in two-dimensional Minkowski spacetime
\cite{FordParker2},
one may suspect that there would be IR
divergences in linearised
gravity in de\ Sitter spacetime as well.  However, it was shown that there 
are no physical IR divergences in the graviton two-point function in
de\ Sitter spacetime which was obtained by analytic continuation from
that on the 4-sphere \cite{Allen86:4}. 
It was also found that the behaviour of mode functions responsible for the 
apparent IR divergences in spatially-flat
de\ Sitter spacetime 
is a gauge artifact \cite{Higuchi87:1}, and it was
shown explicitly that the IR divergences in the graviton two-point function 
in a physical gauge can be gauged away \cite{Allen87:1}.
Thus, it has been established
that there are no physical IR divergences in linearised gravity in 
de\ Sitter spacetime.  However, there is another apparent problem which
is closely related: the graviton two-point function grows logarithmically
with distance.  
If this behaviour were physical, then
it would have a significant effect on physics of inflationary cosmologies.

The aim of this paper is to show that this logarithmic growth
is also a gauge artifact. Specifically,  
we show that the large-distance logarithmic behaviour of the physical graviton
two-point function computed by Allen \cite{Allen87:1}
can be gauged away by  
demonstrating that the same behaviour arises in
the two-point function of pure-gauge form.  
The rest of the paper is organised as follows.  In section 2 we present
the graviton two-point function in a physical gauge
and show that its logarithmically growing
part can be gauged away.  In section 3 we show that this logarithmic behaviour
can be reproduced by a pure-gauge field obeying a relativistic field equation.
We summarise our results and make
some remarks in section 4.  In Appendix A, we list some integrals used in this
work.  Appendix B contains details of the calculations in section 3.
We adopt the metric signature
$(-+++)$ and set $\hbar = c = 16\pi G = 1$
throughout this paper.

\section{The physical graviton two-point function}
\label{secAllen}
We work with the metric which covers half of de\ Sitter spacetime:
\begin{equation}
ds^2=\frac{1}{H^2 \lambda^2}(-d\lambda^2+dx^2+dy^2+dz^2)\,,
\label{metric2}
\end{equation}
where $H^2 = \Lambda/3$, i.e.
$g_{ab} = (H\lambda)^{-2}{\rm diag}(-1,1,1,1)$.
In the expanding half of de\ Sitter spacetime, the parameter $\lambda$ takes
positive values and decreases from $\infty$ to $0$
towards the future.
By letting $g_{ab} = g^{(0)}_{ab} + h_{ab}$, where
$g^{(0)}_{ab}$ is the de\ Sitter metric (\ref{metric2}), and linearising
the Hilbert-Einstein Lagrangian density, we have for linearised gravity
\begin {eqnarray} 
\mathcal{L}& =& \sqrt{-g^{(0)}}
\left[ \frac{1}{2}\nabla_{a}h^{ac}\nabla^{b}h_{bc}
-\frac{1}{4}\nabla_{a}h_{bc}\nabla^{a}h^{bc}
+\frac{1}{4}(\nabla^{a}h-2\nabla^{b} h^{a}_{\ b})\nabla_{a}h \right.
 \nonumber \\
& & \left. \ \ \ \ \ \ \ \ \ \ \ \ 
-\frac{1}{2}H^2\left(
h_{ab}h^{ab}+\frac{1}{2}h^2\right)\right]\,, \label{Lagden}
\end{eqnarray}
where the covariant derivative $\nabla_a$ is compatible with the
background metric $g^{(0)}_{ab}$.  Indices are raised and lowered by 
$g^{(0)}_{ab}$ and $h$ is the trace of $h_{ab}$.
(We will denote $g_{ab}^{(0)}$ by $g_{ab}$ from now on.)
The Lagrangian density (\ref{Lagden}) yields the following
Euler-Lagrange equation:
\begin{eqnarray}
&&\frac{1}{2}\left( \Box h_{ab}-\nabla_{a}\nabla_{c} h^{c}_{\ b}
- \nabla_{b}\nabla_{c} h^{c}_{\ a}+ \nabla_a\nabla_b h\right)
 \nonumber \\
&& + \frac{1}{2}g_{ab}(\nabla_{c}\nabla_{d} h^{cd}
-\Box h) 
-H^2\left(h_{ab}+\frac{1}{2}g_{ab}h\right) = 0\,, \label{Eulag}
\end{eqnarray} 
where $\Box = \nabla_a \nabla^a$.
Equation (\ref{Eulag}) is invariant under the gauge transformations
\begin{equation}
h_{ab} \to h_{ab} + \nabla_a \Lambda_b + \nabla_b \Lambda_a\,.
\label{gauge-transf}
\end{equation}
We will first fix the gauge completely at the classical level. 

By imposing the gauge condition,
\begin{equation}
\nabla_b h^{ab} - \frac{1}{2}\nabla^a h = 0\,, \label{dedonder}
\end{equation}
we find from (\ref{Eulag})
\begin{equation}
\frac{1}{2}\Box h_{ab} - \frac{1}{4}g_{ab}\Box h
- H^2 \left( h_{ab} + \frac{1}{2}g_{ab} h\right) = 0\,.
\label{intereq}
\end{equation}
The trace $h$ can be gauged away as follows.  By taking the
trace of (\ref{intereq}) we have 
\begin{equation}
(\Box + 6H^2)h =0\,.
\end{equation}
Using this equation, we can replace $h_{ab}$ by a traceless field 
satisfying (\ref{intereq}), 
\begin{equation}
\tilde{h}_{ab} = h_{ab} + \frac{1}{6H^2}\nabla_a\nabla_b h\,,
\end{equation}
which is gauge-equivalent to the original field $h_{ab}$.
Thus, the trace $h$ can be gauged away.

The field $\tilde{h}_{ab}$, which we will denote by $h_{ab}$ from now on,
is transverse-traceless, (i.e. it satisfies
$\nabla^{b}h_{ab} = h^{c}_{\ c} = 0$)
and obeys the following equation: 
\begin{equation}
(\Box-2H^2)h_{ab}=0\,.\label{feq}
\end{equation} 
The general solution of this equation can be found, e.g. in \cite{Higuchi87:1}.
Now, this equation allows solutions which are pure gauge:  the field
\begin{equation}
h^{(\xi)}_{ab} = \nabla_{a}\xi_{b}+ \nabla_{b} \xi_{a}  \label{hAab}
\end{equation}
is transverse-traceless and
satisfies (\ref{feq}) if $\nabla^c \xi_{c} = 0$ and
\begin{equation}
(\Box + 3H^2)\xi_{a} = 0\,.  \label{eqforA}
\end{equation}
This gauge freedom allows us to fix the gauge further and 
in effect we can impose the following gauge conditions
on the field $h_{ab}$: 
\begin{eqnarray}
\mathrm{traceless:} & {h^{c}}_{c}=0\,; \nonumber \\
\mathrm{transverse:} & \nabla_{b}h^{ab}=0\,; \nonumber \\
\mathrm{synchronous:} & t^a h_{ab}=0\,, \nonumber 
\end{eqnarray}
where $t^a$ is the future-pointing unit vector parallel to
$-(\partial/\partial\lambda)^a$. 

Allen considered 
quantisation of linearised gravity in this gauge, which we call
the physical gauge, and computed the
symmetrised two-point function.  Here, we present essentially the same
results for the unsymmetrised two-point function,
$G_{aba'b'}(x,x')=\la 0 | h_{ab}(x)h_{a'b'}(x')|0 \ra$, where the state
$|0\rangle$ is the so-called Euclidean vacuum \cite{GibHawk}.
The unprimed indices refer to the spacetime point $x$, whereas the primed
indices refer to the spacetime point $x'$. 

Following Allen, we define the projection operator $P_{ab} = g_{ab}+t_a t_b$
at point $x$,
which projects tensors onto the flat spatial section of constant $\lambda$.
The tensor $P_{ab}$ is 
the metric tensor on this spatial section.  In our coordinate system
it has the form
\begin{equation}
P_{ab} = (H\lambda)^{-2}{\rm diag}(0,1,1,1)\,.
\end{equation}
We define $P_{a'b'}$ to be the same projection operator at point $x'$.
We define a bi-covector $P_{ab'}$ by
\begin{equation}
P_{ab'} = (H^2\lambda\lambda')^{-1} {\rm diag}(0,1,1,1)\,.
\end{equation}
Next we define the comoving spatial separation $r$ of two points
$x=(\lambda,x_1,x_2,x_3)$ and $x'=(\lambda',x'_1, x'_2, x'_3)$ as
\begin{equation}
r(x,x') = \sqrt{(x_1-x'_1)^2 + (x_2-x'_2)^2 + (x_3-x'_3)^2}\,.
\end{equation}
For any given two points $x$ and $x'$ the vectors $V^{a}$ and $V^{a'}$ are 
defined in components as
\begin{equation}
V^a = \frac{H\lambda}{r}(0,{\bf x}-{\bf x}')\,,\ \ \
V^{a'} = \frac{H\lambda'}{r}(0,{\bf x}'-{\bf x})\,.
\end{equation}
The vector $V^a$ and $V^{a'}$ are
the unit vectors at points $x$ and $x'$, respectively, 
which is parallel to the
projection of the tangent vector
to the geodesic joining two points $x$ and $x'$ onto the constant $\lambda$
hypersurface.

The field $h_{ab}(\lambda,{\bf x})$ has the following mode expansion
\cite{FordParker}:
\begin{equation}
h_{ab}(\lambda,{\bf x}) = 
\int d^3{\bf k} \sum_{s=1}^2 \left[
b^{(s)}({\bf k})\frac{H}{4\sqrt{2}\pi} \lambda^{3/2}
H^{(1)}_{3/2}(k\lambda)
\hat{H}_{ab}^{({\bf k},s)} e^{i{\bf k}\cdot{\bf x}} + {\rm h.c.}\right]\,,
\label{modex}
\end{equation}
where the symmetric traceless tensors $\hat{H}_{ab}^{({\bf k},s)}$
satisfy $\hat{H}_{ab}^{({\bf k},s)}\hat{H}^{({\bf k},s')ab} = \delta^{ss'}$ 
and $k^a \hat{H}_{ab}^{({\bf k},s)} = t^a \hat{H}_{ab}^{({\bf k},s)} = 0$.
We have defined $k = \|{\bf k}\|$. Here, 
$d^3{\bf k} = dk_1 dk_2 dk_3$ and
${\bf k}\cdot{\bf x} = k_1x_1+ k_2x_2+ k_3 x_3$.
The Hankel function $H^{(1)}_{3/2}(z)$ is given by
\begin{equation}
H^{(1)}_{3/2}(z) = \sqrt{\frac{2}{\pi}}\left(
-\frac{i}{z^{3/2}} - \frac{1}{z^{1/2}}\right) e^{iz}\,.
\end{equation}
The operators $b^{(s)}({\bf k})$ and  
$b^{(s)}({\bf k})^{\dagger}$ satisfy the commutation relations
\begin{equation}
[ b^{(s)}({\bf k}), b^{(s')}({\bf k}')^\dagger] 
= \delta^{ss'}\delta^3({\bf k}-{\bf k}')\,,
\end{equation}
with all other commutators vanishing.
The Euclidean vacuum $|0\rangle$ is defined by
$b^{(s)}({\bf k})|0\rangle = 0$ for all ${\bf k}$ and $s$.
By using the mode expansion (\ref{modex}) and remembering 
that the two-point function is a maximally symmetric bi-tensor in the spatial 
sections, we find 
\begin{equation}
G_{aba'b'}(x,x') =  
f_{1}(\lambda,\lambda',r) \theta^{(1)}_{aba'b'} + f_{2}(\lambda,\lambda',r) 
{\theta^{(2)}}_{aba'b'} + f_{3}(\lambda,\lambda',r) {\theta^{(3)}}_{aba'b'}
\label{phys2pt}
\end{equation}
where the bi-tensors ${\theta^{(i)}}_{aba'b'} $ are given by
\begin{eqnarray}
{\theta^{(1)}}_{aba'b'}&=& (V_a V_b - \frac{1}{3}P_{ab})(V_{a'} V_{b'} - 
\frac{1}{3}P_{a'b'}) \\
{\theta^{(2)}}_{aba'b'}&=& P_{aa'}P_{bb'}+P_{ba'}P_{ab'}-
\frac{2}{3}P_{ab}P_{a'b'} \\
{\theta^{(3)}}_{aba'b'}&= & 4V_a V_b V_{a'} 
V_{b'}+P_{aa'}V_{b}V_{b'}+P_{ba'}V_a 
V_{b'} +P_{ab'}V_b V_{a'} + P_{bb'}V_a V_{a'}\,. 
\end{eqnarray}
The functions $f_1$, $f_2$ and $f_3$ are
given by
\begin{eqnarray}
f_1 & = & \frac{H^2}{8\pi^2}\left[
\frac{3}{4}V^2(V^2 - 3)\psi_2 - \frac{1}{5}\left( 15V^4 - 40V^2 - 12\right)
\right] \nonumber \\
&& + \frac{H^2\lambda\lambda'}{8\pi^2 r^2}
\left[ \frac{3}{4}(5V^2 - 9) \psi_2 + \frac{15V^4-37V^2 + 16}{1-V^2}\right]
\,, \\
f_2 & = & \frac{H^2}{8\pi^2}\left[
- \frac{1}{5}\psi_1 + \frac{1}{20}V^2(V^2+5)\psi_2 - \frac{1}{75}
(15V^4 + 80V^2 - 32) \right] \nonumber \\
&& + \frac{H^2\lambda\lambda'}{8\pi^2 r^2}
\left[ \frac{1}{4}(V^2+3)\psi_2 
+ \frac{3V^4+7V^2-4}{3(1-V^2)}\right]\,,\\
f_3 & = & \frac{H^2}{8\pi^2}\left[
\frac{1}{4}V^2(V^2+1)\psi_2 - \frac{1}{15}(15V^4+20V^2+8)\right]
\nonumber \\
&& + \frac{H^2\lambda\lambda'}{8\pi^2 r^2}\left[
\frac{1}{4}(5V^2+3)\psi_2 + \frac{15V^4-V^2-8}{3(1-V^2)}\right]\,,
\end{eqnarray}
where
\begin{equation}
V= \frac{\lambda-\lambda'+ i\epsilon}{r}
\end{equation}
and
\begin{eqnarray}
\psi_1 & = & \log[\alpha^4 r^4(1-V^2)^2] + 4\gamma\,, \\
\psi_2 & = & V\log \left( \frac{V+1}{V-1}\right)^2\,.
\end{eqnarray}
Here, $\gamma$ is Euler's constant and
$\alpha (>0)$ is an infrared cut-off.  Since these results are essentially
the same as those in Ref.\ \cite{Allen87:1}
--- recall that we have set $16\pi G = 1$ ---
we omit the details of the
calculations.  The method is similar to that used in the next section.

In the large-$r$ limit these functions become
\begin{eqnarray}
f_{1} & = & \frac{3H^2}{10\pi^2}+O(r^{-2})\,, \label{f11}\\
f_{2} & = & \frac{H^2}{40\pi^2}\left(\frac{32}{15}-2
\log \alpha^2r^2-
4\gamma\right) +O(r^{-2})\,,\label{f22}\\
f_{3} & = & - \frac{H^2}{15\pi^2}+O(r^{-2})\,. \label{f33} 
\end{eqnarray}
Notice that the two-point function $G_{aba'b'}$
is IR divergent and grows
logarithmically with distance due to the behaviour of the function $f_2$.
However, this behaviour can be a
gauge artifact because linearised
gravity has gauge invariance (\ref{gauge-transf}).
The two-point function $G_{aba'b'}$ is physically equivalent to
$G_{aba'b'}^{\rm mod}$ if
\begin{equation} 
G_{aba'b'}(x,x') = 
G_{aba'b'}^{\rm mod}(x,x')
+ \nabla_a \nabla_{a'}T_{bb'} + \nabla_b \nabla_{a'}T_{ab'}
+ \nabla_a \nabla_{b'}T_{ba'} + \nabla_b \nabla_{b'}T_{aa'}\,.
\label{eqn:expansion}
\end{equation}
If there is a field $T_{aa'}$ such that
$\nabla_a \nabla_{a'}T_{bb'} + \nabla_b \nabla_{a'}T_{ab'}
+ \nabla_a \nabla_{b'}T_{ba'} + \nabla_b \nabla_{b'}T_{aa'}$ contains
a term proportional to 
$-(H^2/20\pi^2)\log\alpha^2 r^2 \times \theta^{(2)}_{aba'b'}$,
then the modified two-point function $G^{\rm mod}_{aba'b'}$ will be
IR finite and has no logarithmic growth with distance $r$.  Allen found that
the IR divergence could be gauged away in a similar manner.
However, it is not difficult to show that the logarithmic growth
with distance $r$ 
can be gauged away together with the IR divergence.
Indeed we find
\begin{eqnarray}
\log \alpha^2r^2 \times\theta^{(2)}_{aba'b'}
& = & -\frac{1}{6}\left(\nabla_{a}\nabla_{a'}K_{bb'}
+ \nabla_a \nabla_{b'}K_{ba'} + \nabla_{b}\nabla_{a'}K_{ab'}
+ \nabla_b \nabla_{b'}K_{aa'}\right) \nonumber \\
& & + \frac{1}{3}\left( 
P_{aa'} V_b V_{b'} + P_{a'b}V_a V_{b'}
+ P_{a'b}V_{a'}V_b + P_{bb'}V_a V_{a'} \right.\nonumber \\
&& \ \ \ \ \  
+4P_{ab}V_{a'}V_{b'} + 4P_{a'b'}V_a V_b - 4P_{aa'}P_{bb'} - 4P_{ab'}P_{a'b}
\nonumber \\
&& \ \ \ \ \ \left.
- 8 V_a V_b V_{a'} V_{b'} \right)  
\end{eqnarray}
with
\begin{equation}
K_{aa'} = \frac{r^2}{H^2\lambda\lambda'}
(V_a V_{a'} + 2 P_{aa'})\log \alpha^2 r^2\,.
\end{equation}
Therefore we have
\begin{eqnarray}
G_{aba'b'}(x,x') & = & G_{aba'b'}^{\rm mod}(x,x') \nonumber \\
&&+ \frac{H^2}{120\pi^2}\left( \nabla_a \nabla_{a'}K_{bb'}
+ \nabla_b \nabla_{a'} K_{ab'} + \nabla_a \nabla_{b'}K_{ba'}
+ \nabla_b \nabla_{b'}K_{aa'}\right)\,, 
\end{eqnarray}
where $G_{aba'b'}^{\rm mod}(x,x')$ does not grow logarithmically 
as the function of 
the distance between the two points $x$ and $x'$ and is IR finite. 

This proves that the $\log r$ behaviour of the two-point function
$G_{aba'b'}$
is a gauge artifact.  However, it may be desirable to use 
the two-point function 
$\la 0 | \nabla_{(a}\xi_{b)}(x)\nabla_{(a'}\xi_{b')}(x')| 0 \ra$ of
a vector field $\xi_a$ for gauging away the $\log \alpha^2 r^2$ term
in order to have a better understanding of the logarithmic
growth. 
In the next section we show how this can be done.

\section{Pure-gauge two-point function}

Recall that the tensor $h^{(\xi)}_{ab} = \nabla_a \xi_b + \nabla_b \xi_a$ 
given by (\ref{hAab}) 
satisfies $\nabla^b h^{(\xi)}_{ab} = h^{(\xi)a}_a = 0$ and
$(\Box - 2H^2)h^{(\xi)}_{ab} = 0$.  These are the equations satisfied
by $h_{ab}$ in the physical gauge.  Hence, it is natural to expect that
the two-point function of $h^{(\xi)}_{ab}$ has a structure similar to 
$G_{aba'b'}$.  We will find that this is indeed the case and that
the $\log \alpha^2 r^2$ term in the two-point function
$G_{aba'b'}$ can be gauged away, in the manner described at the end of the
previous section, using the two-point function of
$h^{(\xi)}_{ab}$ with an additional condition $t^a \xi_a = 0$. 

First we note that 
the transverse solutions to equation (\ref{eqforA}) satisfying the 
condition $t^{a}\xi_a = 0$ are 
\cite{Higuchi87:1} 
\begin{equation} 
\xi^{(s)}_{a}({\bf k}, {\bf x}, \lambda) 
=\frac{H}{4\sqrt{2} \pi} \lam^{3/2} H_{5/2}^{(1)}(k \lam) 
e^{i \mathbf{k}\cdot\mathbf{x}}
\hat{h}_{a}^{({\bf k},s)}\,, \label{xic}
\end{equation} 
where the polarisation vectors $\hat{h}_a^{({\bf k},s)}$, $s=1,2$, 
are orthogonal to $k^a$ and $t^a$ and satisfy
\begin{equation}
P^{ab}h_a^{({\bf k},s)}h_b^{({\bf k},s')}=\delta^{ss'}\,.
\end{equation}
The Hankel function $H_{5/2}^{(1)}(z)$ is given by
\begin{equation}
H_{5/2}^{(1)}(z) = \sqrt{\frac{2}{\pi}}\left(
-\frac{3i}{z^{5/2}} - \frac{3}{z^{3/2}} + \frac{i}{z^{1/2}}\right)
e^{iz}\,. \label{Hank5}
\end{equation}

We expand the field $\xi_a(x)$ as
\begin{equation}
\xi_a({\bf x}, \lambda) =
\sum_{s=1}^2 \int d^3{\bf k}  \left[
c^{(s)}({\bf k}) \xi_a^{(s)}({\bf k}, {\bf x},\lambda)
+ c^{(s)}({\bf k})^{\dagger} \overline{\xi_a^{(s)}({\bf k}, {\bf x},\lambda)}
\right]\,.
\end{equation}
We then {\it impose} the commutation relations 
\begin{equation}
[ c^{(s)}({\bf k}), c^{(s')}({\bf k}')^\dagger ]
= \delta^{ss'}\delta^3({\bf k}-{\bf k}')\,,
\end{equation}
with all other commutators being zero. (Since the field $\xi_a$ is a 
fictitious field introduced to rewrite the two-point function $G_{aba'b'}$,
there is no need to {\it derive} these commutation relations.)
Requiring that $a^{(s)}({\bf k})|0\rangle = 0$ for all $s$ and ${\bf k}$,  
we have the two-point function of $\xi_a$,
\begin{eqnarray}
M_{aa'}(x,x')
 & = & \langle 0|\xi_a(x)\xi_{a'}(x')|o\rangle \nonumber \\
& = & 
\int d^3\mathbf{k}\sum_{s=1}^{2}\xi_{a}^{(s)}(\mathbf{k},\mathbf{x},\lambda)
\overline{\xi_{a'}^{(s)}(\mathbf{k},\mathbf{x'},\lambda')}\,.
\label{eqone}
\end{eqnarray}
By using (\ref{xic}) and (\ref{Hank5}), we find 
\begin{eqnarray}
&&\sum_{s=1}^{2} \xi^{(s)}_a ({\bf k},{\bf x},\lambda) 
\overline{\xi^{(s)}_{a'}({\bf k},{\bf x}',\lambda')} \nonumber \\
&& = \frac{H^2 E(k,\lambda,\lambda')}{16\pi^3\lambda\lambda^{\prime}}
e^{ik(\lam-\lam')}
e^{i{\bf k}\cdot({\bf x} - {\bf x}')}
\sum_{s=1}^{2}\hat{h}^{(s)}_{a}( \mathbf{k } ) 
\hat{h}^{(s)}_{a'}(\mathbf{k})\,,
\end{eqnarray}
where
\begin{eqnarray}
E(k,\lambda,\lambda') & = & 
9k^{-5} 
-9i(\lam-\lam')k^{-4} + [9 \lam \lam'-3(\lam^2+{\lam'}^2)]k^{-3}
\nonumber \\ 
&& 
-3i \lam \lam'(\lam-\lam')k^{-2} + \lam^2 {\lam'}^2\,k^{-1}\,.
\end{eqnarray}
Using the properties of $\hat{h}^{(s)}_{a}(\mathbf{k})$, we get 
\begin{equation}
\sum_{s=1}^{2}\hat{h}^{(s)}_{a}( \mathbf{k} ) \hat{h}^{(s)}_{a'}(\mathbf{k})=
\frac{1}{H^2\lam\lam'}\left(
\tildelta_{aa'}-\frac{\tilde{k}_{a}\tilde{k}_{a'}}{k^2}\right)\,,
\label{eqtwo}
\end{equation}
where $\tildelta_{aa'}= \lambda\lambda'P_{aa'}$. 
(In components, $\tildelta_{11}=\tildelta_{22}=\tildelta_{33} = 1$ and
all other components vanish.)
We have also defined the spatial part of $k_a$ as $\tilk_a = P_a^{\ b}k_b$.
It is also convenient to define 
the spacelike projections of the partial derivatives as
$P_a^{\ c}\partial_c = \tilpar_a$ and 
$P_{a'}^{\ c'}\partial_{c'} = \tilpar_{a'}$.
Combining equations (\ref{eqone})--(\ref{eqtwo}), we obtain
\begin{eqnarray}
M_{aa'}(x,x') & = & 
\frac{1}{16\pi^3\lambda^2\lambda^{\prime 2}}\int d^3\mathbf{k} 
\,E(k,\lambda,\lambda')
e^{ik[(\lam-\lam')+i\epsilon]}
e^{i{\bf k}\cdot({\bf x} - {\bf x}')}
\left( \tildelta_{aa'}-\frac{\tilde{k}_{a}\tilde{k}_{a'}}{k^2} 
\right) \nonumber  \\
& = & 
\frac{1}{16\pi^3\lambda^2\lambda^{\prime 2}} \int dk\,k^2 
E(k,\lambda,\lambda')
e^{ik\Dl} \nonumber \\
&& \ \ \ \ \ \ \ \ \ \ \ \ \ \times
\int d\Omega_{\bf k}\,e^{i{\bf k}\cdot({\bf x}-{\bf x}')}
\left( \tildelta_{aa'}-\frac{\tilde{k}_{a}\tilde{k}_{a'}}{k^2}
\right) \,,
\end{eqnarray}
where the convergence factor $e^{-k\epsilon}$ has been introduced to make the 
integral
converge for large wave numbers $k$. We have also defined
$\Dl = \lam-\lam'+i\epsilon$, 
and $\Omega_{\bf k}$ is the solid angle in the ${\bf k}$
space. 
Noting that
$\tilpar_{a}e^{i\mathbf{k}\cdot(\mathbf{x}-\mathbf{x}')}=
i\tilk_{a} e^{i\mathbf{k}\cdot
(\bf{x}-\bf{x}')}$ and
\begin{equation}
\int d\Omega_{\bf k} e^{i\mathbf{k}\cdot(\mathbf{x}-\mathbf{x}')}
= \frac{4\pi\sin kr}{kr}\,, \label{kckc}
\end{equation}
we obtain
\begin{eqnarray}
M_{aa'}(x,x') & = & 
\frac{1}{4\pi^2\lambda^2\lambda^{\prime 2}} \int dk\,k^2 E(k,\lambda,\lambda')
e^{ik\Dl} \nonumber \\
&& \ \ \ \ \
\times \left( \tildelta_{aa'}-\frac{\tilpar_{a} \tilpar_{a'}}{k^2} \right)
\frac{\sin kr}{kr}\,.
\label{xi-expansion}
\end{eqnarray}

We will show that the pure-gauge two-point function, 
\begin{equation}
T_{aba'b'}\equiv 4\la 0| \nabla_{(a}\xi_{b)} \nabla_{(a'}\xi_{b')}|0\ra\,,
\end{equation}
has the same $\log \alpha^2 r^2$ term as $G_{aba'b'}$.
We first note that
\begin{equation}
\nabla_a \xi_b + \nabla_b \xi_a =\mathcal{L}_{\xi}g_{ab} 
=\xi^{c} \partial_{c} g_{ab}+g_{ac}\partial_{b} \xi^{c} +g_{bc} \partial_{a} 
\xi^{c}\,,
\end{equation}
where $\mathcal{L}_\xi$ is the Lie derivative with respect to the vector
$\xi^a$.  
Since the vector field we consider (i.e., $\xi^a$) satisfies $t^a\xi_a=0$,
we have $\xi^{c} \partial_{c}g_{ab}=0$. [Recall that 
the metric $g_{ab}$ depends only on time $\lambda$ and that $t^a$ is 
proportional to $-(\partial/\partial \lambda)^a$.]
Hence we find
\begin{eqnarray}
T_{aba'b'} & \equiv & 4 \la 0 |
\nabla_{(a}\xi_{b)}(x)\nabla_{(a'}\xi_{b')}(x')| 0 \ra  \nonumber \\
& = & \nabla_a \nabla_{a'}M_{bb'} + \nabla_a\nabla_{b'}M_{ba'}
+ \nabla_b \nabla_{a'}M_{ab'} + \nabla_b\nabla_{b'}M_{aa'}\nonumber \\
& = & 
 g_{ac}g_{a'c'}\partial_{b} \partial_{b'} \la 0 | \xi^c \xi^{c'} | 0 \ra 
+ g_{ac}g_{b'c'}\partial_{b} \partial_{a'} \la 0 | \xi^c \xi^{c'} | 0 \ra 
\nonumber \\
&&  + g_{bc}g_{a'c'}\partial_{a} \partial_{b'} \la 0 | \xi^c \xi^{c'} | 0 \ra 
+ g_{bc}g_{b'c'}\partial_{a} \partial_{a'} \la 0 | \xi^c \xi^{c'} | 0 \ra\,. 
\label{gauge}
\end{eqnarray} 

Define $T^{\rm A}_{aba'b'}$ to be the spacelike projection of $T_{aba'b'}$,
\begin{equation}
T^{\rm A}_{aba'b'} = P_{a}^{\ c}P_b^{\ d}P_{a'}^{\ c'}P_{b'}^{\ d'}
T_{cdc'd'}\,.
\end{equation}
Then, 
remembering that $t_a\xi^a = 0$, we find
\begin{equation}
T^{\rm A}_{aba'b'}  = 
\tilpar_{b} \tilpar_{b'} M_{aa'}(x,x')
+ \tilpar_{b}\tilpar_{a'}M_{ab'}(x,x') 
+\tilpar_a\tilpar_{b'}M_{ba'}(x,x')
+\tilpar_{a}\tilpar_{a'}M_{bb'}(x,x')\,,
\label{eqn:gauge}
\end{equation}
where $M_{aa'}$ is given by (\ref{xi-expansion}).  
The components
with one timelike index can be represented by
\begin{eqnarray}
T^{\rm B}_{aa'b'} & \equiv & 
P_{a}^{\ c}t^d P_{a'}^{\ c'}P_{b'}^{\ d'}
T_{cdc'd'}\nonumber \\
& = &
-  \frac{H}{\lambda}\tilpar_{b'}\frac{\partial\ }{\partial \lambda}
[\lambda^2 M_{aa'}(x,x')]
- \frac{H}{\lambda}\tilpar_{a'}\frac{\partial\ }{\partial\lambda}
[ \lambda^2 M_{ab'}(x,x')]\,. \label{one}
\end{eqnarray}
(The minus sign arises due to the fact 
that the parameter $\lambda$ decreases towards the future.)
We define $T^{\rm B}_{aba'}$ as the tensor obtained by interchanging
the primed and unprimed indices in $T^{\rm B}_{aa'b'}$.
The components with two timelike indices can be represented by
\begin{equation}
T^{\rm C}_{ab'}  \equiv  
P_{a}^{\ c}t^d t^{c'}P_{b'}^{\ d'}
T_{cdc'd'}
 = 
\frac{H^2}{\lambda\lambda'}
\frac{\partial^2\ }{\partial \lambda\partial \lambda'}
[\lambda^2 \lambda^{\prime 2} M_{ab'}(x,x')]\,. \label{two}
\end{equation}
The pure-gauge two-point function $T_{aba'b'}$ is given by
\begin{eqnarray}
T_{aba'b'} & = & T^{\rm A}_{aba'b'} + t_b T^{\rm B}_{aa'b'}
+ t_a T^{\rm B}_{ba'b'} + t_{b'}T^{\rm B}_{aba'} 
+ t_{a'}T^{\rm B}_{abb'}  \nonumber \\
&& + t_a t_{a'}T^{\rm C}_{bb'} + t_b t_a'T^{\rm C}_{ab'}
+ t_b t_{a'}T^{\rm C}_{ab'} + t_b t_{b'}T^{\rm C}_{aa'}\,. 
\label{twop}
\end{eqnarray}
We will find that $T^{\rm A}_{aba'b'}$ contains a 
term proportional to 
$\log \alpha^2 r^2\times \theta^{(2)}_{aba'b'}$,
whereas the other two tensors $T^{\rm B}_{aa'b'}$ and
$T^{\rm C}_{ab'}$ contain no terms which grow with the distance $r$.

By using (\ref{xi-expansion}) and (\ref{eqn:gauge}) we have
\begin{eqnarray}
T^{\rm A}_{aba'b'} & = & \frac{1}{4 \pi^2\lambda^2\lambda^{\prime 2}}\int dk\,
k^2 E(k,\lambda,\lambda') 
e^{ik\Dl} 
\nonumber \\
&& \ \ \ \ \ \ \ \times (
\tilpar_{b}\tilpar_{b'}\tildelta_{aa'} 
+ \tilpar_{b}\tilpar_{a'}\tildelta_{ab'}
+\tilpar_{a}\tilpar_{b'}\tildelta_{ba'}
+\tilpar_{a}\tilpar_{a'}\tildelta_{bb'} 
-\frac{4}{k^2}\tilpar_{a}\tilpar_{a'}\tilpar_{b} 
\tilpar_{b'}) \nonumber \\
&& \ \ \ \ \ \ \ \times \frac{\sin kr}{kr}\,.
\label{gauge2}
\end{eqnarray}
This tensor can be evaluated by using
the integration formulas in Appendix A.
The result is
\begin{equation}
T^A_{aba'b'}= T^{(1)}\times \theta^{(1)}_{aba'b'} +
T^{(2)}\times\theta^{(2)}_{aba'b'}+
T^{(3)}\times\theta^{(3)}_{aba'b'}\,,
\end{equation}
where the bi-tensors
$\theta^{(i)}_{aba'b'}$ are defined in section 2,
while the $T^{(i)}$ are given as follows:
\begin{eqnarray}
T^{(1)}&=&-\frac{H^4}{4\pi^2}\left\{\frac{36}{5}+3V^2(4+3V^2(4-
\psi))\right. \nonumber \\
&& \ \ \ \ \ \  +3\frac{\lam\lam'}{r^2}\left[ -\frac{12V^2}{1-V^2}
+8+15V^2(4-\psi)\right]  \nonumber \\
& & \ \ \ \ \ \ \left. 
+3\frac{\lam^2{\lam'}^2}{r^4}\left( 15(4-\psi)+\frac{4(-7+5V^2)}{(1-V^2)^2}
\right)  
\right\}\,,
 \\
T^{(2)}&=&-\frac{H^4}{4\pi^2}\left\{\frac{18}{5}\gamma-
\frac{108}{25}+\frac{1}{5}V^2[4+3V^2(4-\psi)] \right. \nonumber \\
&& \ \ \ 
+\frac{\lam\lam'}{r^2}\left[-2+3V^2(4-\psi)-\frac{6V^2}{1-V^2}\right] 
 \nonumber\\
& &  \ \ \ \left.
+\frac{\lam^2{\lam'}^2}{r^4}\left[ \frac{4(V^2-2)}{(1-V^2)^2}+3(4-
\psi)\right] 
+\frac{9}{5}\log \alpha^2 r^2(1-V^2)\right\}\,,\\
T^{(3)}&=&-\frac{H^4}{4\pi^2}\left\{-\frac{3}{5}+V^2(4+3V^2(4-\psi))-
\frac{3V^2(1+V^2(V^2-2))}{(1-V^2)^3}\right. \nonumber \\
& & \ \ \ 
+\frac{\lam\lam'}{r^2}\left[-1+15V^2(4-\psi)
-\frac{3[-1+V^2(11+V^2(7V^2-17))]}{(1-V^2)^3}\right]\nonumber \\
&& \ \ \ \left.
+\frac{\lam^2{\lam'}^2}{r^4}\left[
15(4-\psi)-\frac{4[9+V^2(5V^2-12)]}{(1-V^2)^3}
\right]\right\}\,. \label{eqn:cut1b}
\end{eqnarray}

To compute $T^{\rm B}_{aa'b'}$ we use (\ref{xi-expansion}) and 
(\ref{one}) and find
\begin{eqnarray}
T^{\rm B}_{aa'b'} & = &
- \frac{H}{4 \pi^2\lambda\lambda^{\prime 2}}
\int dk\,k^2\, \frac{\partial\ }{\partial \lambda}\left[ 
E(k,\lambda,\lambda')
e^{ik\Dl}
\right] \nonumber \\
& & \ \ \ \ \ \ \times
\left( \tildelta_{aa'}\tilpar_{b'} + \tildelta_{ab'}\tilpar_{a'}
-2\frac{\tilpar_{b'}\tilpar_{a} 
\tilpar_{a'}}{k^2}
\right)\frac{\sin kr}{kr}\,. \label{eqn:cut2a}
\end{eqnarray}
Then, using the integral formulas in Appendix A, we have
\begin{equation}
T^{\rm B}_{aa'b'}
=\tilde{T}^{(1)}\beta^{(1)}_{aa'b'}+\tilde{T}^{(2)}\beta^{(2)}_{aa'b'}
\label{resone}
\end{equation}
with
\begin{eqnarray}
\beta^{(1)}_{aa'b'}=P_{aa'}V_{b'}+P_{ab'}V_{a'}
+ 2V_a V_{a'} V_{b'}\,,
 \\
\beta^{(2)}_{aa'b'}= P_{a'b'}V_a
-P_{aa'}V_{b'}
- P_{ab'}V_{a'} - 5V_a V_{a'} V_{b'}\,,
\end{eqnarray}
and
\begin{eqnarray}
\tilde{T}^{(1)} & = & 
\frac{H^4}{4\pi^2}\left\{\frac{3\lam}{r}
\frac{1}{1-V^2}
+\frac{1}{r^3}\frac{2\lam\lam'(3\lam-\lam')}{(1-V^2)^2}
\right. \nonumber \\
&& \ \ \ \ \ 
\left. +\frac{1}{r^4}\frac{8\lam^2{\lam'}^2V}{(1-V^2)^3}
\right\}\,,\\
\tilde{T}^{(2)} & = & \frac{H^4}{4\pi^2}\left\{\frac{2\lam}{r}+\frac{1}{r
^3}\left[\frac{2\lam
\lam'(-3\lam+\lam')}{1-V^2}+6\lam^2(\lam+\lam')\right] \right. \nonumber
\\
&& \ \ \ \ \ \ \ \ \ \left.  
+\frac{1}{r^4}\left[
\frac{2\lam^2{\lam'}^2V(3V^2-5)}{(1-V^2)^2}-\frac{3}{2}\lam^4\psi_{2}\right] 
\right\}\,. \label{eqn:cut2b}
\end{eqnarray}
We have defined $\psi_{2}=V^{-1}\psi $. 

Next we examine the tensor $T^{\rm C}_{ab'}$ by using
(\ref{two}) and (\ref{xi-expansion}).  We first find
\begin{eqnarray}
T^C_{ab'} &  = &
\frac{H^2}{4 \pi^2\lambda\lambda'}\int dk\,k^2\,
\frac{\partial^2\ }{\partial\lambda\partial\lambda'}E(k,\lambda,\lambda')
e^{ik\Dl} \nonumber \\
&& \ \ \ \ \ \times
\left( \tildelta_{ab'}-\frac{\tilpar_{a} \tilpar_{b'}}{k^2} \right)
\frac{\sin kr}{kr}\,.
\label{eqn:cut3a}
\end{eqnarray}
By integrating over $k$, we have
\begin{equation}
T^C_{ab'}=S^{(1)}P_{ab'}+S^{(2)}V_a V_{a'}\,, \label{restwo}
\end{equation}
where the $ S^{(1)}$ and $S^{(2)}$ are given by 
\begin{eqnarray}
S^{(1)}&=&-\frac{\lam\lam' H^4}{2 \pi^2}
\left[\frac{1}{r^2}\frac{V}{(1-V^2)^2}\right.\nonumber \\
&&\ \ \ \ \ \ \ \ \ \ \left. +\frac{1}{r^4}\left(
\frac{2\lam\lam'V^2}{(1-V^2)^2}
+\frac{4\lam\lam'V^2}{(1-V^2)^3}\right) 
\right]\,,
 \\
S^{(2)}&=&-\frac{\lam\lam' H^4}{2 \pi^2}
\left[\frac{1}{r^2}\frac{1}{(1-V^2)^2}
+\frac{1}{r^4}\frac{4\lam\lam'}{(1-V^2)^3} \right]\,.
\label{eqn:cut3b}
\end{eqnarray}
Details of the calculations presented here can be found in Appendix B.

\section{Conclusion}

It can be seen from the results of the previous section
that $T^{B}_{aa'b'}$ and $T^C_{ab'}$ are of order $r^{-1}$ and
of order $r^{-2}$, respectively, and that they are both IR finite.
Hence, we have from (\ref{twop})
\begin{eqnarray}
T_{aba'b'} & =  & T^{\rm A}_{aba'b'} + O(r^{-1})\nonumber \\
& = & -\frac{H^4}{4\pi^2}\left(
\frac{18}{5}\gamma-\frac{108}{25}+\frac{9}{5}\log \alpha^2 r^2
\right)\times \theta^{(2)}_{aba'b'} \nonumber \\
&& 
 -\frac{36}{5} \times 
\theta^{(1)}_{aba'b'}
+\frac{3}{5}\times 
\theta^{(3)}_{aba'b'} + O(r^{-1})\,.
\end{eqnarray}
By comparing this result and 
the physical two-point function $G_{aba'b'}$ given by 
(\ref{phys2pt}) with (\ref{f11})--(\ref{f33}), we find
\begin{equation}
G_{aba'b'}(x,x') = G_{aba'b'}^{\rm mod}(x,x')
+ \frac{4}{9H^2}
\la 0 | \nabla_{(a}\xi_{b)}(x)\nabla_{(a'}\xi_{b')}(x')| 0 \ra\,,
\end{equation}
where the two-point function $G_{aba'b'}^{\rm mod}(x,x')$, which
is related to $G_{aba'b'}(x,x')$ by a gauge transformation, is IR finite
and does not grow
logarithmically with $r$.  Thus, we have shown that the logarithmic behaviour
in $G_{aba'b'}(x,x')$ can be gauged away by a two-point function
of a pure-gauge field. 

Recently, Hawking, Hertog and Turok \cite{Hawketal} have found 
that the physical graviton two-point function is
well-behaved for large distances in open de\ Sitter spacetime.
Now, gauge-invariant correlation functions must be the same in the de\ Sitter
invariant vacuum, whether we use the spatially-flat or open coordinate
system.  Hence,  
their result implies that there cannot be any physical effects due to 
the logarithmic term in the two-point function (\ref{phys2pt}).  This is 
consistent with our result.  Finally, it will be interesting to investigate
the implication of our result for  
two-point functions in covariant gauges
(see, e.g. Ref.\ \cite{Allen86:4,many}).  

\begin{flushleft}
\large{\bf Acknowledgement}
\end{flushleft}
We thank Bernard Kay for useful discussions.
Some of the calculations in this paper were performed using Maple V Release
5.1.
  
\newpage

\section*{Appendix A. Some useful integrals}
\renewcommand{\theequation}{A\arabic{equation}}
\setcounter{equation}{0}
We first obtain
\begin{equation}
\int_{\alpha}^\infty \frac{dk}{k}\, e^{ik(x+i\epsilon)}
 = 
\int_{\alpha (x+i\epsilon)}^\infty \frac{d\kappa}{\kappa}\cos \kappa
+ i \int_{\alpha x}^\infty \frac{d\kappa}{\kappa}\sin \kappa
= - \gamma - \log \alpha (x+i\epsilon) + \frac{\pi i}{2}\,, \label{A1}
\end{equation}
where the terms of order $\alpha$ or higher are neglected.
By differentiating this formula with respect to $x$ we find
\begin{equation}
\int_{0}^\infty dk\,e^{ik(x+i\epsilon)} = \frac{i}{x+i\epsilon}\,. 
\label{A0}
\end{equation}
\begin{equation}
\int_{0}^\infty dk\,k^2 e^{ik(x+i\epsilon)} = \frac{2i}{(x+i\epsilon)^3}\,,
\label{Am2}
\end{equation}
\begin{equation}
\int_{0}^\infty dk\,ke^{ik(x+i\epsilon)} = -\frac{1}{(x+i\epsilon)^2}\,,
\label{Am1}
\end{equation}
We define $K = \frac{\pi i}{2} - \gamma$ and use integration by parts to
find the following formulas:
\begin{equation}
\int_{\alpha}^\infty \frac{dk}{k^2}\,e^{ikx} = 
\frac{1}{\alpha} + ix[K+1 -\log\alpha x]\,, 
\end{equation}
\begin{equation}
\int_{\alpha}^\infty \frac{dk}{k^3}\,e^{ikx} = 
\frac{1}{2\alpha^2}+\frac{ix}{\alpha} -\frac{x^2}{2}\left[
K+\frac{3}{2} -\log\alpha x\right]\,, 
\end{equation}
\begin{equation}
\int_{\alpha}^\infty \frac{dk}{k^4}\,e^{ikx} = 
\frac{1}{3\alpha^3}+\frac{ix}{2\alpha^2} -\frac{x^2}{2\alpha}
-\frac{ix^3}{6}\left[
K+\frac{11}{6} -\log\alpha x\right]\,, 
\end{equation}
\begin{equation}
\int_{\alpha}^\infty \frac{dk}{k^5}\,e^{ikx} = 
\frac{1}{4\alpha^4}+\frac{ix}{3\alpha^3} -\frac{x^2}{4\alpha^2}
-\frac{ix^3}{6\alpha} +\frac{x^4}{24}\left[
K+\frac{25}{12} -\log\alpha x\right]\,, 
\end{equation}
\begin{equation}
\int_{\alpha}^\infty \frac{dk}{k^6}\,e^{ikx} = 
\frac{1}{5\alpha^5}+\frac{ix}{4\alpha^4} -\frac{x^2}{6\alpha^3}
-\frac{ix^3}{12\alpha^2} +\frac{x^4}{24\alpha}-\frac{ix^5}{120}\left[
K+\frac{137}{60} -\log\alpha x\right]\,. 
\end{equation}

\section*{Appendix B. Details of the calculation of the pure-gauge
two-point function}
\renewcommand{\theequation}{B\arabic{equation}}
\setcounter{equation}{0}

\noindent
\large
\textbf{B1. Spatial components}
\normalsize

To calculate $T^{\rm A}_{aba'b'}$ we start from the quantity 
\begin{eqnarray}
A_{aba'b'} &=& \frac{4\pi}{(\lambda\lambda')^2}
\left(\tildelta_{aa'}\tilpar_b \tilpar_{b'}
+\tildelta_{ba'}\tilpar_a \tilpar_{b'}
+\tildelta_{ab'}\tilpar_b \tilpar_{a'}
+\tildelta_{bb'}\tilpar_a \tilpar_{a'}\right. \nonumber \\
&& \ \ \ \ \ \ \ \ \ \ \ \ 
\left. - \frac{4}{k^2}\tilpar_a\tilpar_b \tilpar_{a'}\tilpar_{b'}\right)
\frac{\sin kr}{kr}
\end{eqnarray}
in equation (\ref{gauge2}).
By using 
\begin{equation} 
\tilpar_{a} f(r)=\frac{\partial f}{\partial r} \frac{r_{a}}{r}\,,\ \
\tilpar_{a}r_{a'}=-\tildelta_{aa'}\,,
\end{equation}
and using symmetries of $A_{aba'b'}$ we find
\begin{eqnarray}
A_{aba'b'} & =& \frac{1}{(\lambda\lambda')^{2}}
\left[A^{(1)}\frac{r_{a}r_{b}r_{a'}r_{b'}}{r^4}\right. \nonumber \\
&& \ \ \ \ \ \ \ \ \ \ \ \ +A^{(2)}
\frac{\tildelta_{aa'}r_{b}r_{b'}+\tildelta_{ba'}r_{a}r_{b'}+
\tildelta_{ab'}r_{b}r_{a'}+\tildelta_{bb'}r_{a}r_{a'}}{r^2}  \nonumber \\
&& \ \ \ \ \ \ \ \ \ \ \ \ \left.  +A^{(3)}
\frac{\tildelta_{a'b'}r_{a}r_{b}+\tildelta_{ab}r_{a'}r_{b'}}{r^2}
+A^{(4)} (\tildelta_{a'b} 
\tildelta_{ab'}+\tildelta_{bb'}\tildelta_{aa'}) \right. \nonumber \\
&& \ \ \ \ \ \ \ \ \ \ \ \ 
\left. +A^{(5)}\tildelta_{ab}\tildelta_{a'b'}\right]\,,
\end{eqnarray}
where
\begin{eqnarray}
A^{(1)}&=&4\pi\left(-\frac{4k\sin kr}{r}- 
\frac{40\cos kr}{r^2}+\frac{180\sin kr}{kr^3}
+\frac{420\cos kr}{k^2 r^4}-\frac{420\sin kr }{k^3 r^5} \right)\,, \\
A^{(2)}&=&4\pi\left(- \frac{k\sin kr }{r}-
\frac{7\cos kr}{r^2}+\frac{27\sin kr}{kr^3}
+\frac{60\cos kr}{k^2 r^4}-\frac{60\sin kr}{k^3 r^5} \right)\,, \\
A^{(3)}&=& 4\pi \left(\frac{4\cos kr}{r^2}-\frac{24\sin kr}{kr^3}
-\frac{60\cos kr}{k^2 r^4}+\frac{60\sin kr}{k^3 r^5} \right)\,,  \\
A^{(4)}&=& 4\pi\left(-\frac{2\cos kr}{r^2}+\frac{6\sin kr}{k r^3}
+\frac{12\cos kr}{k^2 r^4}-\frac{12\sin kr}{k^3 r^5} \right)\,, \\
A^{(5)}&=& 4\pi \left( \frac{4\sin kr}{k r^3}+\frac{12\cos kr}{k^2 r^4}
-\frac{12\sin kr}{k^3 r^5} \right)\,.
\end{eqnarray}
Using the definitions for $P_{ab}$ and $V_{a}$, we can express
$A_{aba'b'}$ 
in terms of $P_{ab}$ and $V_{a}$ as
\begin{eqnarray}
H^{-4}A_{aba'b'} & =& A^{(1)}\times V_{a}V_{b}V_{a'}V_{b'} 
 +A^{(2)}\times \left(P_{aa'}V_{b}V_{b'}+P_{ba'}V_{a}V_{b'}+
P_{ab}V_{b}V_{a'}+P_{bb'}V_{a}V_{a'}\right) \nonumber \\
&&+A^{(3)}\times \left(P_{a'b'}V_{a}V_{b}+P_{ab}V_{a'}V_{b'} \right)
+A^{(4)} \times \left(P_{a'b} P_{ab'}+P_{bb'}P_{aa'} \right) \nonumber \\
&& +A^{(5)}\times P_{ab}P_{a'b'}\,.
\end{eqnarray}
The identity 
$\tildelta^{ab} A_{aba'b'}=0$ implies that
$A^{(1)}-4A^{(2)}+3A^{(3)}
= A^{(3)}+2A^{(4)}+3A^{(5)}=0$. 
These equations are indeed satisfied.  They allow us to eliminate
$A^{(1)}$ and $A^{(5)}$.  Thus, we obtain
\begin{equation}
A_{aba'b'}=-3{A}^{(3)}\times \theta^{(1)}_{aba'b'} 
+{A}^{(4)}\times \theta^{(2)}_{aba'b'} 
+{A}^{(2)}\times\theta^{(3)}_{aba'b'}\,, 
\end{equation}
where the $\theta^{(i)}$ are defined in section \ref{secAllen}.
By substituting this in equation (\ref{gauge2}) and
noting that
\begin{eqnarray}
\frac{\cos kr}{k^n}e^{ik\Dl}=\frac{1}{2}\left[\frac{e^{ik(r+\Dl)}}{k^n}
+\frac{e^{ik(-r+\Dl)}}{k^n}\right]\,,\\
\frac{\sin kr}{k^n}e^{ik\Dl}=\frac{1}{2i}\left[\frac{e^{ik(r+\Dl)}}{k^n}
-\frac{e^{ik(-r+\Dl)}}{k^n}\right]\,, 
\end{eqnarray}
we can calculate $T^{\rm A}_{aba'b'}$, using the integrals 
in Appendix A. 
Thus we obtain
\begin{equation}
T^{\rm A}_{aba'b'}= T^{(1)}\times \theta^{(1)}_{aba'b'} +
T^{(2)}\times\theta^{(2)}_{aba'b'}+
T^{(3)}\times\theta^{(3)}_{aba'b'}
\end{equation}
where
\begin{eqnarray}
T^{(1)}&=&-\frac{H^4}{4\pi^2}\left\{
-\frac{36\lam\lam'(\Dl)^2}{r^2(r-\Dl)(r+\Dl)}
-\frac{60\lam^2{\lam'}^2(\Dl)^2}{r^2(r-\Dl)^2 (r+\Dl)^2} \right. \nonumber \\
&& \ \ \ \ \ \ \ \  -\frac{84\lam^2{\lam'}^2}{(r-\Dl)^2 (r+\Dl)^2} 
+\frac{36}{5}
+\frac{36h(\lam,\lam')}{r^4}+\frac{12(\lam^2+{\lam'}^2)}{r^2}\nonumber \\
&& \ \ \ \ \ \ \ \ \left.
-\frac{18(\lam^5-{\lam'}^5)}{r^5}\log\frac{r+\Dl}{-r+\Dl}\right\}\,, \\
T^{(2)}&=&- \frac{H^4}{4\pi^2}\left\{
\frac{18}{5}\gamma -\frac{108}{25}-\frac{2\lam 
\lam'(3 
(\Dl)^2+2
\lam\lam')}{r^2 (r-\Dl)(r+\Dl)}-\frac{4\lam^2{\lam'}^2}{(r-\Dl)^2(r+\Dl)^2} 
\right. \nonumber \\
&& \ \ \ \ \ \ \ \ +\frac{4}{5}\frac{(\Dl)^2}{r^2}+\frac{8\lam\lam'}{r^2}
+\frac{12}{5}\frac{h(\lam,\lam')}{r^4} \nonumber \\
&&\ \ \ \ \ \ \ \  \left. +\frac{9}{5}\log{\alpha^2(r^2-
(\Dl)^2)}
-\frac{6}{5}\frac{\lam^5-{\lam'}^5}{r^5}\log\frac{r+\Dl}{-r+\Dl}\right\}\,,\\
T^{(3)}&=&-\frac{H^4}{4\pi^2}\left\{[3((\Dl)^2-\lam\lam')r^6
+(33\lam\lam'(\Dl)^2-
6(\Dl)^4+36\lam^2{\lam'}^2)r^4 \right. \nonumber \\
&& \ \ \ \ \ \ \ \ +(-51\lam\lam'(\Dl)^4-48\lam^2{\lam'}^2
+3(\Dl)^6)r^2\nonumber \\
&& \ \ \ \ \ \ \ \  +20\lam^2{\lam'}^2(\Dl)^4
+21\lam\lam'(\Dl)^6]/r^2(r+\Dl)^3(r-\Dl)^3 -\frac{3}{5}\nonumber \\ 
&& \ \ \ \ \ \ \ \ \left. +\frac{-\lam\lam'+4(\Dl)^2}{r^2}
+\frac{12h(\lam,\lam')}{r^4}-\frac{6(\lam^5-
{\lam'}^5)}{r^5}\log\frac{r+\Dl}{-
r+\Dl}\right\}\,,
\end{eqnarray}
with
\begin{equation}
h(\lambda,\lambda') = \lambda^4 + \lambda^3\lambda'
+ \lambda^2 {\lambda'}^2 + \lambda {\lambda'}^3 + {\lambda'}^4\,.
\end{equation}
The expression (\ref{eqn:cut1b}) follows by
reexpressing these in terms of $V$ and $\psi$.

\noindent
\large
\textbf{B2. Components with one time index}
\normalsize

We first find a simpler form for the expression (\ref{eqn:cut2a}).
We note that
\begin{eqnarray}
&& \frac{4\pi}{H^3\lambda\lambda'^2}
\left( \tildelta_{aa'}\tilpar_{b'}-\frac{1}{k^2}
\tilpar_{b'}\tilpar_{a}
\tilpar_{a'} \right)\frac{\sin kr}{kr}
\nonumber \\
&&  = 4\pi\left(\frac{\cos kr}{r}
-\frac{\sin kr}{k r^2} \right)\times P_{aa'}V_{b'} \nonumber \\
&& +4\pi\left( \frac{\sin kr}{k r^2}+\frac{3\cos kr}
{k^2 r^3}-\frac{3\sin kr}{k^3 
r^4} \right)
\times \left(-P_{aa'}V_{b'}+P_{a'b'}V_{a}-P_{ab'}V_{a'} \right) 
\nonumber \\
&& 
+4\pi \left( \frac{\cos kr}{r}-\frac{6\sin kr}{k r^2}-\frac{15\cos kr}{k^2
r^3}+\frac{15\sin kr}{k^3 r^4} \right) V_{a}V_{a'}V_{b'}\,.
\end{eqnarray}
We also find 
\begin{equation}
k^2\frac{\partial\ }{\partial \lambda}\left[ E(k,\lambda,\lambda')
e^{ik(\lam-\lam')}\right]
 = [ 3\lam k^{-1} -3i\lam\Dl 
+ \lam \lam'(-\lam'+3\lam)k + i\lam^2
{\lam'}^2k^2]e^{ik\Dl}\,.
\end{equation}
Therefore we have 
\begin{eqnarray}
&& \frac{H}{\lambda}\frac{\partial\ }{\partial\lambda}\tilpar_{b'}
[\lambda^2 M_{aa'}(x,x')] \nonumber \\
&& = 
\frac{H^4}{16 \pi^3}
\int dk \,
\left\{ 3\lam k^{-1} -3i\lam\Dl
+ \lam \lam' (-\lam'+3\lam)k + i\lam^2 {\lam'}^2k^2
e^{ik\Dl}\right\} \nonumber \\
&& \ \ \ \ \ \ \ 
\times\left\{
4\pi\left[\frac{\cos kr}{r}-\frac{\sin kr}{k r^2}\right] 
\times P_{aa'}V_{b'}\right. \nonumber \\
&& +4\pi\left[\frac{\sin kr}
{k r^2}+\frac{3\cos kr}{k^2 r^3}-\frac{3\sin kr}{k^3 
r^4} \right]
\times \left(-P_{aa'}V_{b'}+P_{a'b'}V_{a}-P_{ab'}V_{a'} \right) 
\nonumber \\
&& \left. +4\pi \left[ \frac{\cos kr}{r}-\frac{6\sin kr}{k r^2}-
\frac{15\cos kr}{k^2
r^3}+\frac{15\sin kr}{k^3 r^4} \right] V_{a}V_{a'}V_{b'}\right\}\,.
\end{eqnarray}
We obtain the analogous expression for
$(H/\lambda)(\partial/\partial\lambda)\tilpar_{a'}
[\lambda^2 M_{ab'}(x,x')]$
by interchanging $a'$ with $b'$.
Thus, we have
\begin{eqnarray}
T^{\rm B}_{aa'b'}&=&
-\frac{H}{\lambda}\frac{\partial\ }{\partial\lambda}\tilpar_{b'}
[\lambda^2 M_{aa'}(x,x')] 
- \frac{H}{\lambda}\frac{\partial\ }{\partial\lambda}\tilpar_{a'}
[\lambda^2 M_{ab'}(x,x')] \nonumber \\ 
&= & -\frac{H^4}{16 \pi^3}
\int dk 
\left[ 3\lam k^{-1} -3i\lam\Dl +
\lam \lam' (-\lam'+3\lam)k + i\lam^2 {\lam'}^2 k^{2}\right]\nonumber \\
&& \ \ \ \ \ \ \ \ \ \ \times e^{ik\Dl}
\times D_{aa'b'}\,,
\end{eqnarray}
where $D_{aa'b'}$ is given by
\begin{equation}
D_{aa'b'}=D^{(1)}\beta^{(1)}_{aa'b'}+D^{(2)}\beta^{(2)}_{aa'b'}
\end{equation}
with
\begin{eqnarray}
D^{(1)}&=&4\pi\left(\frac{\cos kr}{r}-
\frac{\sin kr}{k r^2}\right) \nonumber\\
D^{(2)}&=&8\pi\left( \frac{\sin kr}{k r^2}+\frac{3\cos kr}{k^2 r^3}
-\frac{3\sin kr}{k^3 r^4} \right) 
\end{eqnarray}
and
\begin{eqnarray}
\beta^{(1)}_{aa'b'}=P_{aa'}V_{b'}+P_{ab'}V_{a'}
+ 2V_a V_{a'} V_{b'}\,,
 \\
\beta^{(2)}_{aa'b'}= P_{a'b'}V_a
-P_{aa'}V_{b'}
- P_{ab'}V_{a'} - 5V_a V_{a'} V_{b'}\,.
\end{eqnarray}
Again, by performing the $k$ integration using the formulas in Appendix A
we find the result in (\ref{resone}).

\noindent
\large
\textbf{B3. Components with two time indices}
\normalsize

We begin by noting that 
\begin{eqnarray}
&& \frac{4\pi}{H^2\lam\lam'}
\left( \tildelta_{ab'}-\frac{\tilpar_{a} \tilpar_{b'}}{k^2} 
\right)\frac{\sin kr}{kr}
\nonumber \\
&&=   4\pi\left(\frac{\sin kr}{kr}+\frac{\cos kr}{k^2 r^2}
-\frac{\sin kr}{k^3 r^3} 
\right)  
P_{aa'} \nonumber \\
&& \ \ 
 +4\pi\left(\frac{\sin kr}{kr}+\frac{3\cos kr}
{k^2 r^2}-\frac{3\sin kr}{k^3 r^3} 
\right)
V_{a}V_{a'}
\end{eqnarray}
and
\begin{equation}
k^2\,\frac{\partial^2\ }{\partial\lambda\partial\lambda'}
\left[ E(k,\lambda,\lambda')
e^{ik\Dl}\right]
= \left( \lam\lam' k 
-i\lam\lam' \Dl k^2 + \lam^2 {\lam'}^2 k^3 \right)e^{ik\Dl}\,.
\end{equation}
Therefore we find
\begin{eqnarray}
T^{\rm C}_{ab'}&=&
\frac{H^2}{\lambda\lambda'}\frac{\partial^2\ }{\partial\lam\partial \lam'}
[(\lam\lam')^2 M_{ab'}(x,x')]
\nonumber \\
&=&\frac{H^4}{16 \pi^3}
\int dk 
\left( \lam\lam'k 
-i\lam\lam' \Dl k^2 + \lam^2 {\lam'}^2 k^3\right)e^{ik\Dl}
\times E_{ab'}\,,
\end{eqnarray}
where $E_{ab'}$ is given by
\begin{equation}
E_{ab'}=E^{(1)}P_{ab'}
+E^{(2)}V_a V_{b'}
\end{equation}
with
\begin{eqnarray}
E^{(1)}&=&4\pi\left(\frac{\sin kr}{kr}+\frac{\cos kr}{k^2 r^2}-
\frac{\sin kr}{k^3 r^3} \right)\,,\\
E^{(2)}&=&4\pi\left( \frac{\sin kr}{kr}+\frac{3\cos kr}{k^2 r^2}-
\frac{3\sin kr}{k^3 r^3}
\right)\,. 
\end{eqnarray}
We integrate over $k$ using the formulas in Appendix A and obtain
\begin{equation}
T_{ab'}=S^{(1)}P_{ab'}+S^{(2)}V_a V_{b'}\,,
\end{equation}
with
\begin{eqnarray}
S^{(1)}&=&- \frac{H^4\lam\lam'}{2 \pi^2}
\left[\frac{\lam^2+{\lam'}^2}{(r-
\Dl)^2(r+\Dl)^2}
+\frac{4\lam\lam'(\Dl)^2}{(r+\Dl)^3(r-\Dl)^3} \right]\,,\\ 
S^{(2)}&=& - \frac{H^4\lam\lam'r^2}{2 \pi^2}
\left[\frac{1}{(r-\Dl)^2(r+\Dl)^2}
+\frac{4\lam\lam'}{(r+\Dl)^3(r-\Dl)^3} \right]\,.
\end{eqnarray}
By expressing
these in terms of $V$ and powers of $\lam\lam'$ we find
(\ref{eqn:cut3b}).


\newpage

\begin{thebibliography}{99}

\bibitem{HawkEllis} S.W.\ Hawking and G.F.R.\ Ellis, The large-scale
structure of space-time (Cambridge Univ.\ Press, Cambridge, 1973).
\bibitem{Inflation} A.H.\ Guth, {\it Phys.\ Rev.\ D} {\bf 23}, 347 (1981);\\
A.D.\ Linde, {\it Phys.\ Lett.} {\bf 108B}, 389 (1982);\\
A.\ Albrecht and P.J.\ Steinhardt, {\it Phys.\ Rev.\ Lett.} {\bf 48}, 1220
(1982).
\bibitem{FordParker} L.H.\ Ford and L.\ Parker, {\it Phys.\ Rev.\ D}
{\bf 16}, 1601 (1977).
\bibitem{FordParker2} L.H.\ Ford and L.\ Parker, {\it Phys.\ Rev.\ D}
{\bf 16}, 245 (1977); \\
B.\ Ratra, {\it Phys.\ Rev.\ D} {\bf 31}, 1931 (1985);\\
B.\ Allen, {\it Phys.\ Rev.\ D} {\bf 32}, 3136 (1985). 
\bibitem{Allen86:4} B.\ Allen, 
{\it Phys.\ Rev. 
D} {\bf 34},
3680 (1986).
\bibitem{Higuchi87:1}A.\ Higuchi, 
\emph{Nucl.\ Phys.} \textbf{B282}, 397 (1987).
\bibitem{Allen87:1} B.\ Allen, 
\emph{Nucl.\ Phys.},
\textbf{B287}, 743 (1987).
\bibitem{GibHawk} G.\ Gibbons and S.W.\ Hawking, {\it Phys.\ Rev. D} 15,
2738 (1977). 
\bibitem{Hawketal} S.W.\ Hawking, T.\ Hertog and N.\ Turok, hep-th/0003016.
\bibitem{many} B.\ Allen and M.\ Turyn, {\it Nucl.\ Phys.} {\bf B292},
813 (1987); \\
E.\ Floratos, J.\ Illiopoulos and T.N.\ Tomaras, {\it Phys.\ Lett.} 
{\bf 197B}, 373 (1987);\\
I.\ Antoniadis and E.\ Mottola, {\it J.\ Math.\ Phys.} {\bf 32}, 1037
(1991);\\
N.C.\ Tsamis and R.P.\ Woodard, {\it Commun.\ Math.\ Phys.} {\bf 162},
217 (1994);\\
M.V.\ Takook, gr-qc/0001052.
\end{thebibliography}
\end{document}